\documentclass[pdftex,notwocolumn,epjc3]{svjour3}          

\RequirePackage[T1]{fontenc}

\smartqed  

\RequirePackage{graphicx}
\RequirePackage{mathptmx}      
\RequirePackage{flushend}
\RequirePackage[numbers,sort&compress]{natbib}
\RequirePackage[colorlinks,citecolor=blue,urlcolor=blue,linkcolor=blue]{hyperref}

\journalname{Eur. Phys. J. C}
\usepackage{subfigure}
\usepackage{pstricks}
\usepackage{color}
\usepackage{amsfonts}
\usepackage{mathrsfs}
\usepackage{epsfig}
\usepackage{amsmath,amssymb,graphicx,latexsym}
\DeclareMathOperator{\sech}{sech}
\usepackage{ulem}

\newcommand{\be}{\begin{equation}}
	\newcommand{\ee}{\end{equation}}
\newcommand{\bea}{\begin{eqnarray}}
	\newcommand{\eea}{\end{eqnarray}}
\newcommand{\fig}{Fig.\ref}
\newcommand{\eq}{Eq.\eqref}
\newcommand{\phicmb}{\phi_{\rm CMB}}
\newcommand{\phiend}{\phi_{\rm end}}
\newcommand{\lambert}{{\cal LW}[\alpha, N_{\rm CMB}]}
\newcommand{\lambertn}{{\cal LW}[\alpha, N]}
\def\bea{\begin{eqnarray}}
\def\eea{\end{eqnarray}}
\def\ba{\begin{array}}
\def\ea{\end{array}}

\def\beq{\begin{equation}}
\def\eeq{\end{equation}}

\begin{document}

\title{Mutated hilltop inflation revisited}


\author{Barun Kumar Pal\thanksref{e1,addr1}
}

\thankstext{e1}{e-mail:terminatorbarun@gmail.com}

\institute{Department of Mathematics,  Netaji Nagar College for Women, Kolkata 700 092, INDIA\label{addr1}
}

\date{Received: date / Accepted: date}

\maketitle

\begin{abstract}
In this work we re-investigate  pros and cons of mutated hilltop inflation. Applying Hamilton-Jacobi formalism we solve inflationary dynamics and find that inflation goes on along the ${\cal W}_{-1}$ branch of the Lambert function. Depending on the model parameter mutated hilltop model renders two types of inflationary solutions: one corresponds to small inflaton excursion during observable inflation and the other  describes large field inflation. The inflationary observables from curvature perturbation are in tune with the current data for a wide range of the model parameter. The small field branch predicts negligible amount of tensor to scalar ratio $r\sim \mathcal{O}(10^{-4})$, while the large field sector is capable of generating high amplitude for tensor perturbations, $r\sim \mathcal{O}(10^{-1})$. Also, the spectral index is almost independent of the model parameter along with a very small negative amount of scalar running. Finally we find that the mutated hilltop inflation closely resembles the $\alpha$-attractor class of inflationary models in the limit of $\alpha\phi\gg 1$.
\end{abstract}
\section{Introduction} \label{sec1}
The standard model of hot Big-Bang scenario is instrumental in explaining the nucleosynthesis, expanding universe along with the formation of cosmic microwave background (CMB henceforth). But there are few limitations in the likes of {\it flatness problem, homogeneity problem} etc., which can not be answered within the limit of Big-Bang cosmology. In order to overcome these shortcomings an early phase of accelerated expansion -- {\it cosmic inflation} was proposed \cite{starobinsky1980}, \cite{sato1981}, \cite{guth1981}, \cite{albrecht1982}, \cite{linde1982}. Big-Bang theory is incomplete without inflation and turns into brawny when combined with the paradigm of inflation. Though inflation was initiated to solve the cosmological puzzles, but the most impressive impact of inflation happens to be its ability to provide persuasive mechanism for the origin of cosmological fluctuations observed in the large scale structure and CMB. Nowadays inflation is the best bet for the origin of primordial perturbations. 

Since its inception, almost four decades ago, inflation has remained the most powerful tool to explain the early universe when combined with big-bang scenario. It is still a paradigm due to the elusive nature  of the scalar field(s), {\it inflaton}, responsible for inflation and the unknown shape of the potential involved. That the potential should be sufficiently flat to render almost scale invariant curvature perturbation along with tensor perturbation \cite{starobinsky1979}, \cite{mukhanov1981}, \cite{hawking1982}, \cite{starobinsky1982}, \cite{guth1982}, \cite{mukhanov1990}, \cite{stewart1993} has been only understood so far. As a result there are many inflationary models in the literature.  With the advent of highly precise observational data from various probes \cite{cobe1992},\cite{wmap9:hinshaw2012}, \cite{planck2015param}, \cite{planck2015inf}, the window has become thinner, but still allowing numerous models to pass through \cite{martin2014}, \cite{martin2014best}. The recent  detection  of  astronomical gravity waves by LIGO \cite{ligo2016a}, \cite{ligo2016b} has made the grudging cosmologists waiting for primordial gravity waves which are believed to be produced during inflation through tensor perturbation. The upcoming stage-IV CMB experiments are expected to constrain the inflationary models further \cite{abazajian2016cmb} by detecting primordial gravity waves. 

The most efficient method for studying inflation is the slow-roll approximation \cite{starobinsky1978}, \cite{liddle1994}, where the kinetic energy is assumed to be very small compared to the potential energy. But this is not the only way for successful implementation of inflation and solutions outside slow-roll approximation have been found \cite{kinney1997}. In order to study inflationary paradigm irrespective of slow-roll approximation Hamilton-Jacobi formalism \cite{salopek1990}, \cite{muslimov1990} has turned out to be very handy. Here the inflaton itself is treated as the evolution parameter instead of time, and the Friedmann equation becomes first order which is easy to extract underlying physics from. Another interesting class of inflationary models has  been introduced very recently,  {\it constant-roll inflation} \cite{motohashi2015, motohashi2017a, motohashi2017b}, where the inflaton rolls at a constant rate. 

Here we would like to study single field mutated hilltop model (MHI henceforth) of inflation \cite{barunmhi}, \cite{barunmhip} using Hamilton-Jacobi formalism. In MHI observable inflation occurs as the scalar field rolls down towards the potential minimum.  So MHI does not correspond to usual hilltop inflation \cite{boubekeur2005}, \cite{kohri2007} directly, but the shape of the inflaton potential is somewhat similar to the mutated hilltop in hybrid inflation and hence the name. We shall see that for a wide range of values of the model parameter MHI provides inflationary solution consistent with recent observations. 
Our analysis also reveals that MHI has two different branches of inflationary solutions: one corresponds to small field inflation and the other represents large field inflation. In earlier studies \cite{barunmhi}, \cite{barunmhip} we have reported that MHI can only produce a negligible amount of tensor to scalar ratio, $r\sim 10^{-4}$. But, we shall see here that it is capable of generating $r$ as large as ${\cal O}(10^{-1})$ depending on the model parameter.  Consequently a wide range of $r$, $10^{-4}\lesssim r\lesssim 10^{-1}$,  can be addressed by MHI. Recent data from Planck \cite{planck2015param}, \cite{planck2015inf} , \cite{bicep2016improved}
has reported an upper bound $r_{0.002}<0.07$ and upcoming CMB-S4 experiments are expected to survey tensor to scalar ratio up to $r\sim 2\times10^{-3}$ \cite{abazajian2016cmb}. So sooner or later the model can be tested with the observations. The prediction for inflationary observables from MHI are in tune with recent observations. Further, MHI predicts spectral index which is almost independent of model parameter along with small negative scalar running consistent with current data. It has been  also found that  MHI closely resembles the $\alpha$-attractor class of inflationary models \cite{kallosh2013, galante2015} in the limit of $\alpha\phi\gg 1$.

In Section \ref{hj} we have briefly reviewed Hamilton-Jacobi formalism. In the next Section \ref{mhi} we have discussed about the MHI in Hamilton-Jacobi formalism and have shown resemblance with $\alpha-$Attractor class of inflationary models in Section \ref{alphaa}. Finally we conclude in Section \ref{con}.

\section{Quick Look at  Hamilton Jacobi Formalism }\label{hj}
The Hamilton-Jacobi formalism allows us to recast the Friedmann equation into the following form \cite{salopek1990}, \cite{muslimov1990}, \cite{kinney1997}, \cite{barunquasi}
\bea
\left[H^{'}(\phi)\right]^2 -\frac{3}{2\rm M_{P}^2}
H(\phi)^2&=&-\frac{1}{2\rm M_{P}^4}V(\phi)\label{hamilton}\\
\dot{\phi}&=&-2\rm M_{P}^2 H'(\phi)\label{phidot}
\eea
where prime and dot denote derivatives with respect to the scalar field $\phi$ and time respectively, and ${\rm M_{ P}}\equiv\frac{1}{\sqrt{8\pi G}}$ is the reduced Planck mass. The associated inflationary potential can then be  found by rearranging the terms of Eqn.(\ref{hamilton}) 
\beq\label{potential}
V(\phi)=3\rm M_{P}^2H^2(\phi)\left[1-\frac{1}{3}\epsilon_{_H}\right]
\eeq
where $ \epsilon_{\rm H}$ has been defined as
\beq\label{epsilon} \epsilon_{\rm H}=2\rm M_{P}^2\left(\frac{H^{'}(\phi)}{H(\phi)}
\right)^2.
\eeq 
We further have 
\beq\label{adot}
\frac{\ddot{a}}{a}=H^2(\phi)\left[1-\rm\epsilon_{_H}\right]. 
\eeq
Therefore accelerated expansion takes place when $\epsilon_{\rm H}<1$ and ends exactly at $\epsilon_{\rm H}=1$.
The evolution of the scale factor turns out to be 
\beq\label{scale}
a\propto \exp\left[\int\frac{H}{\dot{\phi}}d\phi\right].
\eeq
The amount of inflation is expressed in terms of number of e-foldings and defined as
\beq\label{efol}
N\equiv\ln\frac{a_{\rm end}}{a}=\frac{1}{\rm M_P}\int_{\phi_{\rm  end}}^{\phi}\frac{1}{\sqrt{2\epsilon_{\rm H}}}d\phi.
\eeq 
We have defined $N$ in such a way that at the end of inflation $N=0$ and $N$ increases as we go back in time. The observable parameters are generally evaluated when there are  $55-65$ e-foldings still left before the end of inflation. 
It is customary to define another parameter by
\begin{equation}\label{eta}
	\eta_{\rm H}=2\rm M_{P}^2~ \frac{H^{''}(\phi)}{H(\phi)}. 
\end{equation}
It is worthwhile to mention here that the parameters $\epsilon_{\rm H}$ and $\eta_{\rm H}$ are not the usual  slow-roll parameters. But in the
slow-roll limit  $\epsilon_{\rm H} \rightarrow\epsilon$ and $\eta_{\rm H}\rightarrow\eta-\epsilon$ \cite{liddle1994}, $\epsilon$ and $\eta$  being usual {\it potential slow-roll} parameters.

Though we have not included higher order slow-roll parameters in the present analysis, another  widely used  higher order slow-roll parameter is defined as follows,
\bea
\zeta_{_H}^2(\phi)&\equiv& 4M_P^4\ \frac{H'(\phi)H'''(\phi)}{H^2(\phi)}
\eea 
In Section \ref{mhi}, we shall present variation of the solution of the equation $\zeta_{_H}(\phi)=1$ 
 with the model parameter.
\section{Mutated Hilltop Inflation: The Model}\label{mhi}
The potential we would like to study has the following form \cite{barunmhi}, \cite{barunmhip} 
\beq\label{mhi_potential}
V(\phi)=V_{0} \left[1-\sech(\alpha\phi)\right]
\eeq 
where $V_0$ is the typical energy scale of inflation and $\alpha$ is a parameter having dimension of inverse Planck mass. The potential under consideration does not actually represent typical hilltop inflation \cite{boubekeur2005}, \cite{kohri2007}, but the form of the potential is somewhat similar to mutated hilltop inflation in hybrid scenario and hence the name. Accelerated expansion takes place as the inflaton rolls towards the potential minimum. Not only that, from \eq{mhi_potential} it is obvious that $V(\phi_{\rm min})=V'(\phi_{\rm min})=0$ which is significantly different from the usual hilltop potential. 

The associated Hubble parameter may be written as
\beq\label{hubble}
H(\phi)\simeq \sqrt{\frac{V_0}{3M_{\rm P}^2}} \left[1-\sech(\alpha\phi)\right]^{\frac{1}{2}}
\eeq 
The value of the constants can be fixed from the conditions for successful inflation and the observational bounds. 

\begin{figure}
	\centerline{\includegraphics[width=12.cm, height=8.cm]{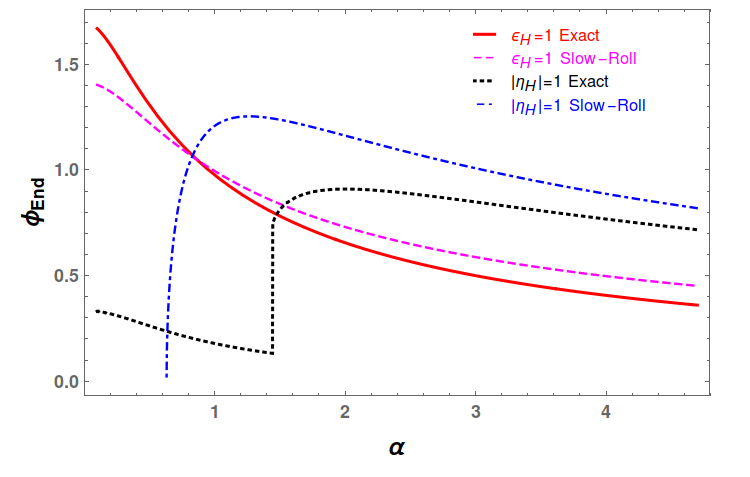}}
	\caption{\label{fig_phiend} The solid red line is the solution of $\epsilon_{_H}=1$ obtained numerically from the background evolution and the dashed magenta line is the solution of $\epsilon_{_H}=1$ using slow-roll approximation. The black dotted line is the solution of $|\eta_{_H}|=1$ obtained numerically from the background evolution  and the dot-dashed blue line is the solution of $|\eta_{_H}|=1$ in the slow-roll approximation.}
\end{figure}

The parameters $\epsilon_{_H}, \ \eta_{_H}, \ \zeta_{_H}$  in the Hamilton-Jacobi formalism now take the form 
\bea\label{slowroll}
	\epsilon_{_H} &\simeq& \frac{\rm M_{P}^2\alpha^2}{2}\frac{\sech^2(\alpha\phi)\tanh^2(\alpha\phi)}{\left(1-\sech(\alpha\phi)\right)^2}\nonumber\\
	\eta_{_H}
	&\simeq&-\frac{\rm M_{P}^2\alpha^2}{2}\sech(\alpha\phi)\left[2+3\sech(\alpha\phi)\right]\nonumber\\
	\zeta_{_H}&\simeq&\frac{M_P^2}{2} \sqrt{\alpha^4 (6 \cosh (\alpha \phi)+2 \cosh (2 \alpha \phi)-13) \coth
		^2\left(\alpha\phi/2\right) \text{sech}^4(\alpha \phi)}
\eea
In \fig{fig_phiend} we have plotted the solutions of $\epsilon_{_H}=1$ and $|{\eta_{_H}}(\phi)|=1$ obtained by solving the background evolution numerically  along with those found using the approximate form \eq{hubble} for the Hubble parameter.  The background evolution was found numerically by solving \eq{hamilton} with the potential given by \eq{mhi_potential}. From the figure it is clear that $|\eta_{_H}|=1$ occurs well before $\epsilon_{_H}=1$ for $\alpha\geq\alpha_{\rm eq}$ in both the cases, where $\alpha_{\rm eq}$  represents the value of $\alpha$ for the simultaneous occurrence of $|\eta_{_H}|=1 \ {\rm and} \ \epsilon_{_H}=1$ (values of $\alpha_{\rm eq}$ are different in exact \& approximate solutions). 

In the Hamilton-Jacobi formalism with Hubble slow-roll parameters  the end of inflation is explicitly given by  $\epsilon_{_H}=1$ at $\phiend$ \cite{liddle1994, kinney1997}, which is also clear from \eq{adot}. From the \fig{fig_phiend} it may be seen that the exact numerical solution of $\epsilon_{_H}=1$ (the solid red line) is  close to that  obtained by using the approximated form of the Hubble parameter  (dashed magenta line), and differs slightly for $|\eta_{_H}|=1$. But as 
we are using an approximated form of the Hubble parameter, we shall consider the end of inflation to be where slow-roll approximation breaks, i.e. $\max_\phi\{\epsilon_{_H}=1, \ |\eta_{_H}|=1\}$. 
Now the solution of   $\epsilon_{_H}=1$ is the root of the following equation 
\beq\label{phiend1}
{{\rm \rm M_{P}^2}}\ \alpha^2\sech^2(\alpha\phi_\epsilon)\left(1-\sech^2(\alpha\phi_\epsilon)\right) \simeq2\left(1-\sech(\alpha\phi_\epsilon)\right)^2.
\eeq 
\eq{phiend1} can be solved analytically and the relevant  solution turns out to be
\bea
\phi_\epsilon &\simeq&{\rm M_P}  b^{-1}\sech^{-1}\frac{1}{3}\left[-1+\frac{b^2-6}{b\left(36b-b^3+3\sqrt{6}\sqrt{4+22b^2-b^4}\right)^{1/3}}\right.\nonumber\\
&+&\left.\frac{\left(36b-b^3+3\sqrt{6}\sqrt{4+22b^2-b^4}\right)^{1/3}}{b}\right]\label{phiend}
\eea
where $b\equiv\alpha{\rm M_P}$. On the other hand  the solution of $|\eta_{_H}|=1$ is given by 
\beq
\phi_\eta\label{etaend} \simeq\alpha^{-1}\sech^{-1}\frac{1}{3}\left[-1+\sqrt{1+\frac{6}{b^2}}\right].
\eeq
So the end of inflation may be written as $\phiend\simeq\max\{\phi_\epsilon, \ \phi_\eta\}$, i.e. $\phiend\simeq \phi_\epsilon$ and $\phiend\simeq \phi_\eta$ for $\alpha\leq\alpha_{\rm eq}$ and $\alpha>\alpha_{\rm eq}$ respectively. 

\begin{figure}
	\centerline{\includegraphics[width=11.cm, height=7.cm]{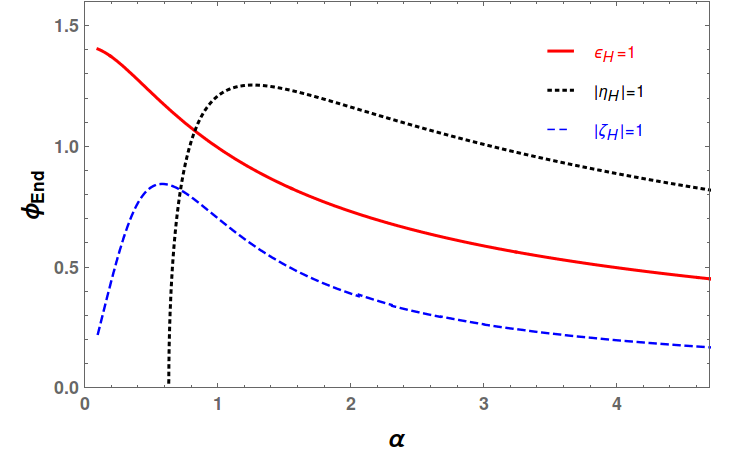}}
	\caption{\label{fig_erp}  The  solutions of  $\epsilon_{\rm H}=1$ (solid red line), $|\rm \eta_{_H}|=1$  (dotted black line) and $|\zeta_{_H}|=1$  (dashed blue line) have been plotted. The approximated form of  the Hubble parameter has been used to find these solutions.}
\end{figure}
In \fig{fig_erp} we have plotted the solutions of $\epsilon_{\rm H}=1$, $|\rm \eta_{_H}|=1$   and $|\zeta_{_H}|=1$ using the Hubble parameter as given by \eq{hubble}. From the figure we see that $|\eta_{_H}|$ becomes order of unity before $\epsilon_{\rm H}$, but $|\zeta_{_H}|$ remains small. 

From now on all the results that we shall present in this article are based on the approximated form of the Hubble parameter and without considering the effect of higher order slow-roll parameters.

\subsection{Number of e-foldings}
The number of e-foldings in MHI is found to have  the following  form
\beq\label{efoldings}
N(\phi)\simeq\frac{1}{\alpha^2\rm \rm M_{P}^2}\left[ \cosh(\alpha\phi)- \cosh(\alpha\phi_{\rm end}) -2 \ln\frac{\cosh(\alpha\phi/2)}{\cosh(\alpha\phi_{\rm end}/2) } \right].
\eeq
The above \eq{efoldings} can be analytically inverted to get the scalar field as a function of e-foldings as follows 
\bea\label{phin}
\phi &\simeq& \alpha^{-1}\cosh^{-1}\left[-1-{\cal W}_{-1}\bigg(-\left[\cosh(\alpha\phiend)+1\right]e^{-\rm M_P^2\alpha^2 N - 1 -\cosh(\alpha\phiend)} \bigg)\right]\nonumber\\
&=& \alpha^{-1}\cosh^{-1}\left({\cal LW}[\alpha,N] \right)
\eea 
where we have defined 
${\cal LW}[\alpha,N]\equiv -1-{\cal W}_{-1}\left(-\left[\cosh(\alpha\phiend)+1\right]e^{-\rm M_P^2\alpha^2 N - 1 -\cosh(\alpha\phiend)} \right)$ and  ${\cal W}_{-1}$ is the Lambert function. From the above \eq{phin}, one can see that mutated hilltop inflation occurs along the ${\cal W}_{-1}$ branch of the Lambert function, first pointed out in Ref.\cite{martin2014}. The value of the inflaton when cosmological scale leaves the horizon, $\phi_{\rm CMB}$, is then given by  
\beq
\phi_{\rm CMB} \simeq \alpha^{-1}\cosh^{-1}\left(\lambert\right)
\eeq 
The slow-roll parameters now can be expressed as a function of the e-foldings
\bea
\epsilon_{_H}&=&\frac{1}{2\rm M_{P}^2}\left(\frac{d\phi}{dN}\right)^2\nonumber\\
&\simeq&\frac{\rm M_P^2\alpha^2}{2}\frac{\lambertn+1}{\left(\lambertn-1\right)\lambertn^2}\\
\eta_{_H}&=&\left(\frac{d^2\phi}{dN^2}\right)\left(\frac{d\phi}{dN}\right)^{-1} +\epsilon_{_H}\label{n_ep}\nonumber\\
&\simeq&-\frac{\rm M_P^2\alpha^2}{2}\frac{2\lambertn+3}{\lambertn^2}\label{n_eta}
\eea
This makes life simpler as now all the inflationary observable parameters when derived in the slow-roll limit can be expressed as a function of $N$. 

\subsection{The Lyth Bound for MHI}
The fluctuations in the tensor modes solely depends on the Hubble parameter whereas curvature perturbation is a function of the Hubble parameter and inflaton. Consequently, tensor to scalar ratio determines excursion of the inflaton during observable inflation, first shown in Ref.\cite{lyth1997} and known as Lyth bound 
\bea\label{lythd}
\Delta\phi&=&\frac{\rm m_P}{8\sqrt{\pi}}\int_0^{N_{\rm CMB}}\sqrt{r}\ dN
\eea
where $\rm m_P={2\sqrt{2\pi}}M_P$ is the actual Planck mass.
$\Delta\phi\geq \ {\rm m_P}$ corresponds to large field model and $\Delta\phi<  \ {\rm m_P}$ small field models. One expects to get larger tensor to scalar ratio, $r$, where $\Delta\phi\geq \ {\rm m_P}$ due to the higher energy scale required for successfully explaining the observable parameters. For the model under consideration we have found 
\bea\label{lyth}
\Delta\phi&\simeq&\alpha^{-1}\cosh^{-1}\left(\lambert\right)- \alpha^{-1}\cosh^{-1}\left({\cal LW}\left[\alpha, 0\right]\right)
\eea
\begin{figure}
	\centerline{\includegraphics[width=11cm, height=7cm]{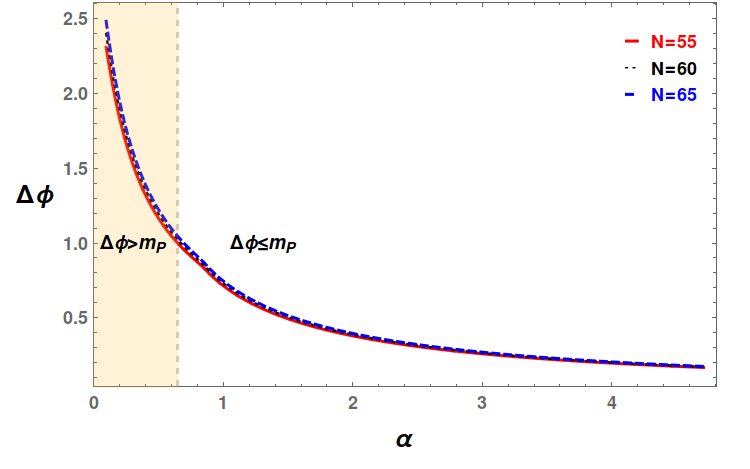}}
	\caption{\label{fig_lyth} Excursion of the scalar field (in the unit of $m_p$) with the model parameter $\alpha$ for three different values of e-foldings, $N=55,60,65$. The dotted vertical line corresponds to the value of $\alpha$ at which $\Delta\phi=\ m_P$ which has been estimated by considering $N_{\rm CMB}=55$.}
\end{figure}
In \fig{fig_lyth} we have shown the variation of the scalar field excursion in the unit of  ${\rm m}_{\rm P}$,  with the model parameter $\alpha$. From the figure it is obvious that the mutated hilltop model of inflation has small excursion of the inflaton for $\alpha\geq \alpha_{\Delta\phi=1}$ and large field excursion for  $\alpha<\alpha_{\Delta\phi=1}$, where $\alpha_{\Delta\phi=1}$ is the solution of \eq{lyth} for $\alpha$ with $\Delta\phi=\ {\rm m_P}$. So this model is capable of addressing both the large and small field inflationary scenarios for suitable values of the model parameter.
\begin{figure}
	\centerline{\includegraphics[width=11.cm, height=7.cm]{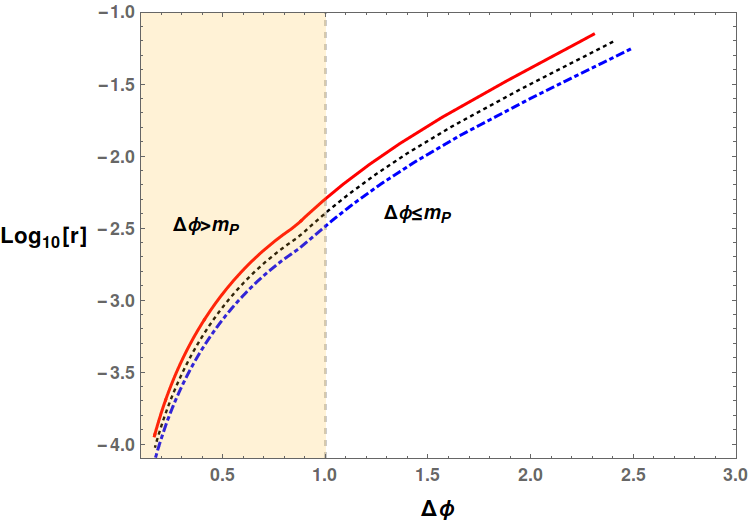}}
	\caption{\label{fig_lythr} The logarithmic variation of the tensor to scalar ratio with $\Delta\phi$ for three values of $N_{\rm CMB}$ has been plotted. Red solid line for $N_{\rm CMB}=55$, black dotted curve represents $N_{\rm CMB}=60$ and the blue dashed line for $N_{\rm CMB}=65$. }
\end{figure}
In \fig{fig_lythr} we have shown the variation of $\rm Log_{10} r$ with $\Delta\phi$. From the figure we see that small field MHI may give rise to negligible amount of tensor to scalar ratio, $r\sim\mathcal{O}(10^{-4})$, on the other hand for large $\Delta\phi$, $r$ can be as large as $\mathcal{O}(10^{-1})$. 


\subsection{Inflationary Observables in the Slow-Roll Limit}
The inflationary observable parameters can be expressed  in terms of the  slow-roll parameters. The power spectrum of the curvature perturbation turns out to be
\bea
P_{\cal R}&\simeq&\frac{1}{16\pi^2 \rm M_{P}^4} \bigg[\frac{H(\phi)^2}{H'(\phi)}\bigg]^2_{\phi=\phicmb}\nonumber\\
&\simeq&\frac{V_0}{12\pi^2\alpha^2 \rm M_{P}^6}\frac{\lambert\left(\lambert-1\right)^2}{\lambert+1}\label{sr_pr}
\eea

\begin{figure}
	\centerline{\includegraphics[width=11.cm, height=7.cm]{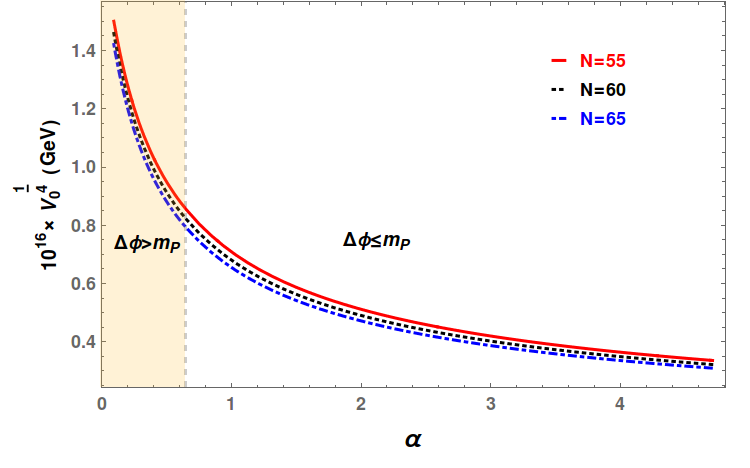}}
	\caption{\label{fig_energy} The energy scale in the unit of GeV for MHI has been plotted with the model parameter $\alpha$ for $N_{\rm CMB}=55,  60,  65$. The shaded region corresponds to the large field sector for MHI.}
\end{figure}
In \fig{fig_energy} we have shown the variation of the typical energy scale associated with MHI for different values of the model parameter. From the figure it is clear that maximum energy scale that can be achieved is ${\mathcal{O}(10^{16})}$ GeV. To determine this energy scale we have used $P_{\mathcal{R}}=2.142\times 10^{-9}$ from Planck 2015 result \cite{planck2015inf}. The scale dependence of the spectrum of curvature perturbation is described by spectral index. In MHI we have found 
\bea
n_{_S} &\simeq& 1-4\epsilon_{_H}|_{\phi=\phicmb}+2\eta_{_H}|_{\phi=\phicmb} \nonumber\\
&\simeq&1-\rm M_{P}^2\alpha^2\frac{2\lambert^2+3\lambert-1}{\lambert^2\left(\lambert-1\right)}\label{sr_ns}
\eea
\begin{figure}
	\centerline{\includegraphics[width=11.cm, height=7.cm]{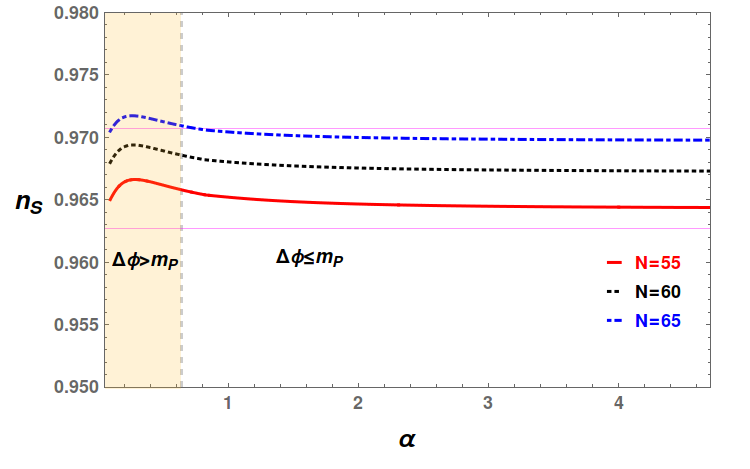}}
	\caption{\label{fig_ns} The variation of the  scalar spectral index with $\alpha$ for three different values of $N_{\rm CMB}$ has been plotted. The shaded vertical region is the result for large field sector of MHI. The two horizontal lines (magenta) are for Planck 2015 upper \& lower bounds on $n_{\rm S}$.}
\end{figure}
In \fig{fig_ns} we have shown how the scalar spectral index changes with the model parameter. We also see that spectral index is almost constant in both the large and small field sector of MHI. The current bound on $n_{_S}$ from Planck 2015 has also been plotted.

The scale dependence of the spectral index itself is estimated from the scalar running and we have
\bea
n_{_S}'&\simeq&-2\rm M{_P}^4 \ H'(\phi)H'''(\phi)/H^2(\phi)|_{\phi=\phicmb} +16\epsilon_{_H}\eta_{_H}|_{\phi=\phicmb} -8\epsilon_{_H}^2|_{\phi=\phicmb}\label{sr_rr}\nonumber\\
&\simeq&-\frac{\rm M{_P}^4\alpha^4}{2}\left[-32+30\lambert+33\cosh(2\cosh^{-1}\lambert)\right.\nonumber\\
&+&\left.\cosh(3\cosh^{-1}\lambert)\right] \lambert^{-4}(\lambert+1)^{-1}\nonumber\\
&\times& (\lambert-1)^{-2}\label{sr_run}
\eea
\begin{figure}
	\centerline{\includegraphics[width=11.cm, height=7.cm]{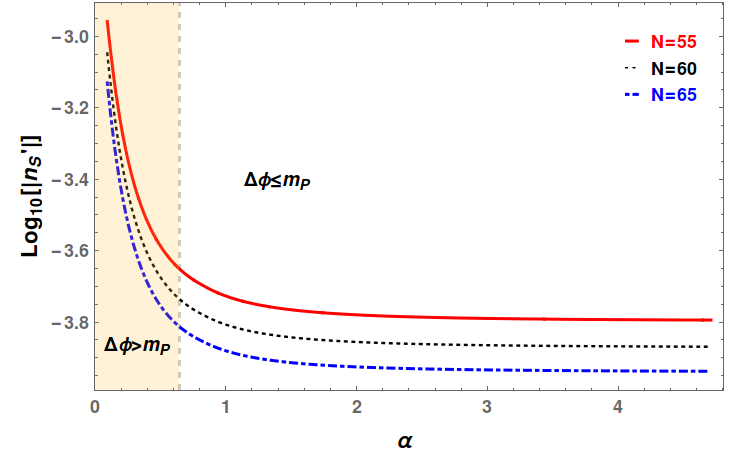}}
	\caption{\label{fig_running}The logarithmic variation of the absolute value of the scalar running with $\alpha$ for three different values of $N_{\rm CMB}$ has been plotted. The shaded vertical region is the result for large field sector of the model under consideration.}
\end{figure}
Here in \fig{fig_running} logarithmic variation of the absolute value of scalar running with $\alpha$ has been plotted.  From the figure it is clear that MHI predicts very small running of the spectral index. The maximum amount of scalar running that can achieved in MHI is  $|n_{\rm S}'|\sim 10^{-3}$. 

Finally, the tensor to scalar ratio is found to have the following form
\bea
r&\simeq&16\epsilon_{_H}|_{\phi=\phicmb}\nonumber\\
&\simeq&8\rm M_{P}^2\alpha^2\frac{\lambert+1}{\lambert^2\left(\lambert-1\right)}\label{sr_r}
\eea
\begin{figure}
	\centerline{\includegraphics[width=11.cm, height=7.cm]{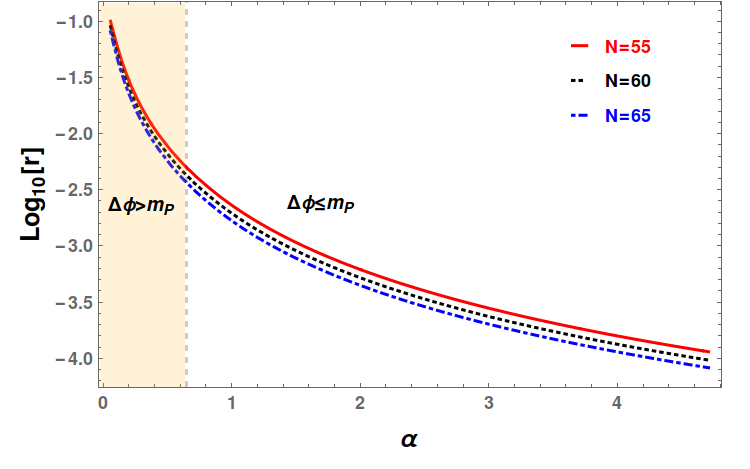}}
	\caption{\label{fig_r} Logarithmic variation of the tensor to scalar ratio with $\alpha$ for three different values of $N_{\rm CMB}$ has been plotted. The shaded vertical region corresponds to the large field sector of the MHI.}
\end{figure}
In \fig{fig_r} we have plotted the tensor to scalar ratio (in ${\rm Log_{10}}$ scale) with $\alpha$ for three different values of $N_{\rm CMB}$. We see that MHI can address wide range of values of tensor to scalar ratio,  $10^{-4}\lesssim r< 10^{-1}$   depending on the model parameter $\alpha$. But $r\gtrsim0.07$ has to be discarded which is observationally forbidden \cite{bicep2016improved}, which effectively determines a rough lower bound on the model parameters, $\alpha>0.094561$.

\begin{figure}
	\centerline{\includegraphics[width=11.cm, height=7.cm]{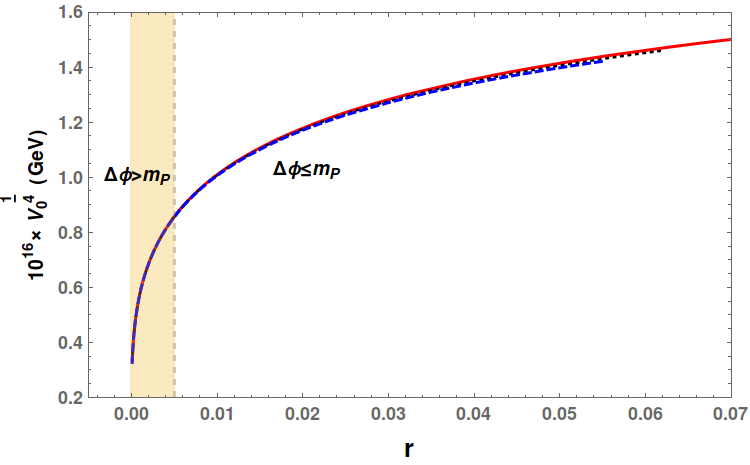}}
	\caption{\label{fig_vor} Variation of the MHI energy scale with  $r$ has been shown for three different values of $N_{\rm CMB}$. Red solid line is for $N_{\rm CMB}=55$, black dotted curve represents $N_{\rm CMB}=60$ and the blue dashed line for $N_{\rm CMB}=65$. The shaded 
		vertical region represents small field sector of MHI.}
\end{figure}
In \fig{fig_vor} we have shown variation of the MHI energy scale with the tensor to scalar ratio. So in order to achieve $r \sim 0.07$ we need an energy scale $V_0^{\frac{1}{4}}\sim 1.5\times10^{16}$ GeV.

\section{Similarity with $\alpha-$Attractor Class of Inflationary Models}\label{alphaa}
Now we shall show the resemblance of MHI with the $\alpha-$attractor class of models \cite{kallosh2013, galante2015} in the limit of  $\alpha\phi\gg1$. 
During inflation, within the slow-roll limit, the spectral index is given by
\bea
n_{_S}-1 &\simeq& -4\epsilon_{_H}+2\eta_{_H}\label{attractor_ns}\nonumber\\
&\simeq& -M_{P}^{2}\alpha^{2}\frac{3\sech^2(\alpha\phi)-\sech^3(\alpha\phi)+2\sech(\alpha\phi)}{1-\sech(\alpha\phi)}\nonumber\\
&\approx&-2M_{P}^{2}\alpha^{2}\sech(\alpha\phi)
\eea
where in the last step we have retained only the leading order term, keeping in mind that we are considering the limit $\alpha\phi\gg1$. Also, the scalar to tensor ratio, $r$, may be expressed as 
\bea
r&\simeq&16\epsilon_{_H}\label{attractor_r}\nonumber\\
&\simeq& 8M_{P}^{2}\alpha^{2}\frac{\sech^2(\alpha\phi)\left[1-\sech^2(\alpha\phi)\right]}{(1-\sech(\alpha\phi))^2}\nonumber\\
&\approx&8M_{P}^{2}\alpha^{2}\sech^2(\alpha\phi)
\eea
where again in the last line we have kept leading order term in $\sech(\alpha\phi)$. 
Now the number of $e-$foldings, $N$, is approximately given by \eq{efoldings}, which in the limit  $\alpha\phi\gg1$   may be written as 
\beq\label{attractor_n}
N\approx M_{P}^{-2}\alpha^{-2} \cosh(\alpha\phi)
\eeq
In order to derive the above \eq{attractor_n} we have made use of the fact that logarithmic function varies slowly and neglected the constant terms $\cosh(\alpha\phi_{\rm end})$ and $\ln[\cosh(\alpha\phi_{\rm end}/2)]$ during inflation. 
Combining \eq{attractor_ns}, \eq{attractor_r} and \eq{attractor_n}, we see that in the limit of $\alpha\phi\gg1$
\beq\label{attractor}
n_{_S}-1\approx-\frac{2}{N} \quad \mbox{and} \quad r\approx \frac{1}{M_{P}^{2}\alpha^{2}}\frac{8}{ N^2}
\eeq
So from \eq{attractor} we see that during inflation when $\alpha\phi\gg1$, MHI indeed belongs to the class of $\alpha-$attractor models.
\begin{figure}
	\centerline{\includegraphics[width=12.cm, height=8.cm]{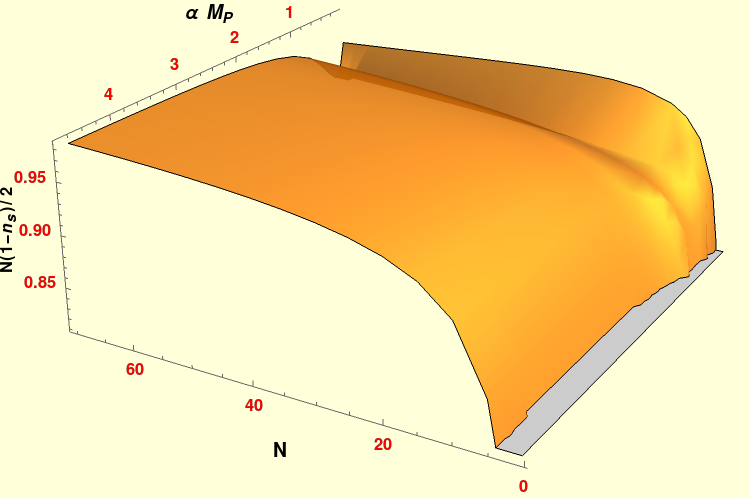}}
	\caption{\label{fig_alphans3d} The variation of $N(1-n_{_S})/2$ in MHI with the model parameter $\alpha$ and the number of $e$-foldings, N, has been shown.}
\end{figure}
\begin{figure}
	\centerline{\includegraphics[width=12.cm, height=8.cm]{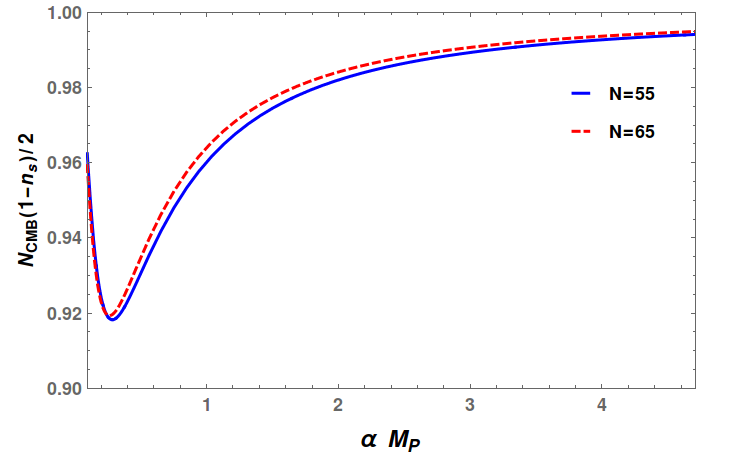}}
	\caption{\label{fig_alphans} Variation of $N_{\rm CMB}(1-n_{_S})/2$ in MHI with $\alpha$ has been shown for two different values of $N_{\rm CMB}$. Blue solid line is for $N_{\rm CMB}=55$, red dotted curve represents $N_{\rm CMB}=65$.}
\end{figure}
In \fig{fig_alphans3d} we have shown the variation of $N(1-n_{_S})/2$ with the model parameter $\alpha$ and  the number of $e$-foldings, N. From the figure it is clear that for large values of the model parameter MHI belongs to the $\alpha-$attractor class of models and deviation occurs for the small values of the model parameter.  In \fig{fig_alphans} we have shown the variation of $N_{\rm CMB}(1-n_{_S})/2$ with the model parameter $\alpha$ for two different values of $N_{\rm CMB}$. 

\begin{figure}
	\centerline{\includegraphics[width=12.cm, height=8.cm]{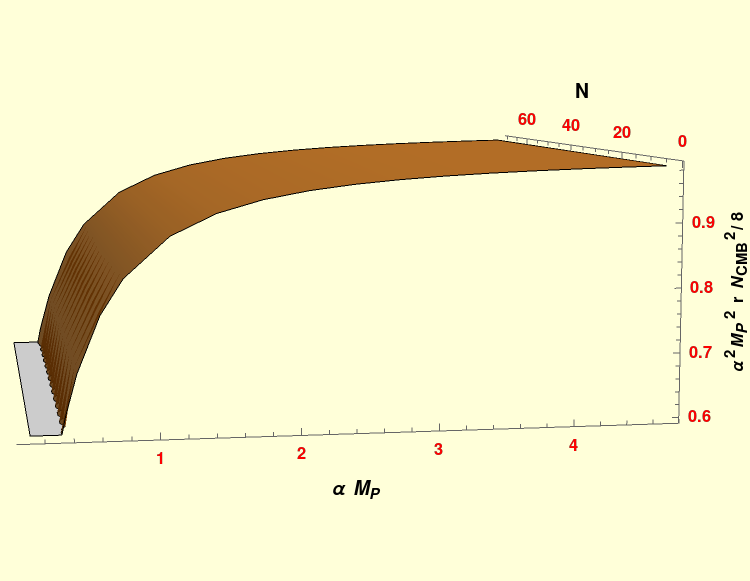}}
	\caption{\label{fig_alphan3d} Variation of $\alpha^2M_P^2\ r\ N^2/8$ in MHI with the model parameter $\alpha$ and the number of $e$-foldings, N, has been shown. }
\end{figure}
\begin{figure}
	\centerline{\includegraphics[width=12.cm, height=8.cm]{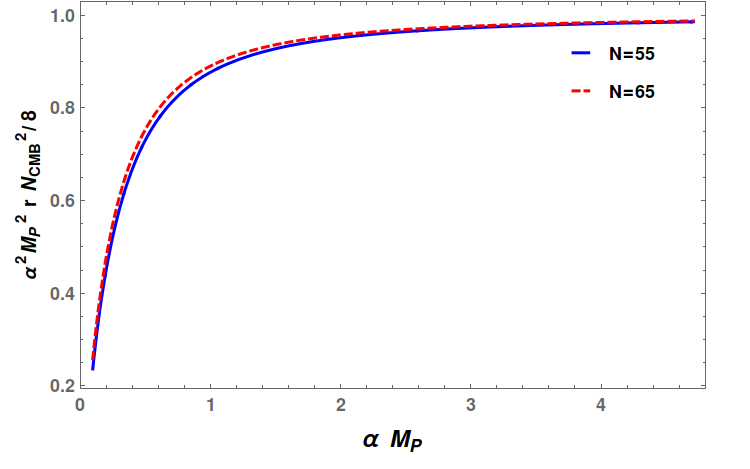}}
	\caption{\label{fig_alphan} Variation of $\alpha^2M_P^2\ r\ N^2_{\rm CMB}/8$ in MHI with $\alpha$ has been shown for three different values of $N_{\rm CMB}$. Blue solid line is for $N_{\rm CMB}=55$, red dotted curve represents $N_{\rm CMB}=65$.}
\end{figure}
In \fig{fig_alphan3d} we have shown the variation of $\alpha^2M_P^2\ r \ N^2/8$ with the model parameter $\alpha$  and the number of $e$-foldings, N. From the figure it is clear that for large values of the model parameter MHI indeed belongs to the $\alpha-$Attractor class of models and small deviation occurs for the small values of the model parameter. In \fig{fig_alphan} we have plotted $\alpha^2M_P^2\ r \ N^2_{\rm CMB}/8$ for two different values of $N_{\rm CMB}$. 

So, from the above results it is quite transparent that MHI indeed falls into the category of $\alpha$-attractor class of inflationary models in the limit of $\alpha\phi\gg1$. 

\section{Conclusion}\label{con}
In this article we have revisited mutated hilltop inflation driven by a hyperbolic potential. Employing Hamilton-Jacobi formulation we found that inflation ends naturally through the violation slow-roll approximation. 
More interestingly, MHI has two different branches of inflationary solution. One corresponds to large field variation  and the other represents small change in inflaton during the observable inflation depending on the model parameter.

Observable parameters as derived from this model are in tune with the latest observations for a wide range of the model parameter, $\alpha{\rm M_P}\gtrsim0.094561$. The scalar spectral index is found to be independent of the model parameter with a small negative running.  We have also found that MHI can address a broad range of the tensor to scalar ratio, $0.0001\lesssim r <0.07$.  In a nutshell, MHI though does not belong to the usual hilltop inflation is extremely attractive with only one model parameter consistent with recent observations. Not only that, in the limit of  $\alpha\phi\gg 1$, MHI closely resembles the $\alpha$-attractor class of inflationary models. 
\section*{Acknowledgment}
I would like to thank Supratik Pal for useful discussions and constructive suggestions. I would also like to thank IUCAA, Pune for giving me the opportunity  to carry on research work through their Associateship Program. Finally author would like to sincerely thank anonymous reviewers for their critical and constructive suggestions on the first version of this work.   

\bibliographystyle{unsrt}
\bibliography{references}

\end{document}